\documentclass[table,xcdraw,conference,a4paper,10pt]{IEEEtran}
\IEEEoverridecommandlockouts

\usepackage{cite}

\usepackage{amsmath,amssymb,amsfonts}
\usepackage{graphicx}
\usepackage{hologo}
\usepackage{threeparttable}
\graphicspath{ {./pic/}}
\usepackage{lscape}
\usepackage{comment}
\usepackage{wasysym}
\usepackage{textcomp}
\usepackage{color}
\usepackage{float}
\usepackage{multirow}
\usepackage{hhline}
\usepackage{subcaption}
\usepackage[font=small]{caption}
\usepackage{array}
\usepackage{tabularx}
\usepackage{tikz}
\usepackage{xcolor}
\usepackage{highlight}
\usepackage{array}
\usepackage{multicol}
\usepackage{collcell}
\usepackage[T1]{fontenc}
\usepackage{enumitem}
\usepackage{algorithm}

\usepackage[utf8]{inputenc}
\usepackage{fourier} 
\usepackage{array}
\usepackage{makecell}
\usepackage[english]{babel}
\usepackage[noend]{algpseudocode}

\algrenewcommand\algorithmicindent{0.6em}%
\usepackage{todonotes}
\usepackage{titlesec}
\usepackage{mathtools}

\usepackage{amsmath}
\usepackage{MnSymbol}
\usepackage{pifont}
 \usepackage{trace}
\usepackage{amssymb}
\usepackage{xcolor}
\usepackage{listings}
\usepackage{listings,color}
\usepackage{tablefootnote}
\usepackage{booktabs}
\definecolor{darkgreen}{rgb}{0,0.5,0}

\usepackage[most]{tcolorbox}
\newcounter{wavenum}

\newcommand*{\clki}{
  \draw (t_cur) -- ++(0,.3) -- ++(.5,0) -- ++(0,-.6) -- ++(.5,0) -- ++(0,.3)
    node[time] (t_cur) {};
}

\newcommand{\nextwave}[1]{
  \path (0,\value{wavenum}) node[left] {#1} node[time] (t_cur) {};
  \addtocounter{wavenum}{-1}
}

\usepackage{setspace}
\newif\ifblind
\blindtrue

\usepackage{tikz}
\usepackage{pgfplots}
\pgfplotsset{compat=1.8}
\newcommand{\cellwidth}{5mm}%
\newcommand{\cellheight}{0.63em}%
\newcommand{\MinNumber}{0.0}%
\newcommand{\MidNumber}{0.5} %
\newcommand{\MaxNumber}{1.0}%

\newcommand{\mybox}[1]{%
\ifdim #1 pt > \MidNumber pt
            \pgfmathsetmacro{\PercentColor}{max(min(100.0*(#1 - \MidNumber)/(\MaxNumber-\MidNumber),100.0),0.00)} %
            \hspace{-0.33em}\colorbox{green!\PercentColor!yellow}{\parbox[][\cellheight][t]{\cellwidth}{\centering \footnotesize #1}}
        \else
            \pgfmathsetmacro{\PercentColor}{max(min(100.0*(\MidNumber - #1)/(\MidNumber-\MinNumber),100.0),0.00)} %
            \hspace{-0.33em}\colorbox{red!\PercentColor!yellow}{\parbox[][\cellheight][t]{\cellwidth}{\centering \footnotesize #1}}
        \fi}

\newcolumntype{R}{>{\collectcell\mybox}c<{\endcollectcell}}

\definecolor{vgreen}{RGB}{104,180,104}
\definecolor{vblue}{RGB}{49,49,255}
\definecolor{vorange}{RGB}{255,143,102}
\definecolor{backcolour}{rgb}{0.95,0.95,0.92}
\lstdefinestyle{verilog-style}
{
    language=Verilog,
    backgroundcolor=\color{backcolour},   
    basicstyle=\scriptsize\ttfamily,
    keywordstyle=\color{vblue},
    identifierstyle=\color{black},
    commentstyle=\color{vgreen},
    numbersep=9pt,
    tabsize=3,
    literate=*{:}{:}1
}

\makeatletter
\newcommand*\@lbracket{[}
\newcommand*\@rbracket{]}
\newcommand*\@colon{:}
\newcommand*\colorIndex{%
    \edef\@temp{\the\lst@token}%
    \ifx\@temp\@lbracket \color{black}%
    \else\ifx\@temp\@rbracket \color{black}%
    \else\ifx\@temp\@colon \color{black}%
    \fi\fi\fi
}
\makeatother

\IEEEoverridecommandlockouts
\begin{document}
\bstctlcite{IEEEexample:BSTcontrol}
\title{Automatic Assertion Mining in Assertion-Based Verification: Techniques, Challenges, and Future Directions
}
\author{\IEEEauthorblockN{Mohammad Reza Heidari Iman, Giorgio Di Natale, and Katell Morin-Allory \\
 Univ. Grenoble Alpes, CNRS, Grenoble INP*\thanks{* Institut National Polytechnique Grenoble Alpes}, TIMA, 38000 Grenoble, France\\
 \{mohammadreza.heidari-iman, giorgio.di-natale, katell.morin-allory\}@univ-grenoble-alpes.fr}
}

\maketitle

\begin{abstract}

Functional verification increasingly relies on Assertion-Based Verification (ABV), which has become a key approach for verifying hardware designs due to its efficiency and effectiveness. Central to ABV are automatic assertion miners, which apply different techniques to generate assertions automatically. This paper reviews the most recent, advanced, and widely adopted assertion miners, offering a comparative analysis of their methodologies. The goal is to provide researchers and verification practitioners with insights into the capabilities and limitations of existing miners. By identifying their shortcomings, this work also points toward directions for developing more powerful and advanced assertion miners in the future.



\end{abstract}

\IEEEpeerreviewmaketitle

\section{Introduction}
\label{sec:intro}
Embedded systems and microelectronic components are employed in a wide range of devices and industries, including automotive and medical sectors, as well as in general safety-critical applications \cite{6384987,Reza}. Ensuring the correctness of the functionality of these systems is crucial, as any failure could lead to disastrous consequences, ranging from financial or ecological harm to, in some cases, death or serious injury \cite{6386877,Reza, Damiano}. One approach to preventing such disasters in embedded systems is functional verification.

Functional verification techniques aim to ensure that a hardware design behaves in accordance with the specified set of requirements outlined in its specification \cite{10.5555/519876, HeidariImanBook2025}. 
As hardware designs continue to grow in size and complexity, verification has become a critical component of modern design flows \cite{articlesocverif, HeidariIman2025chaptverif}. Today, approximately 51\% of the total cost and time in the design process is allocated to verification \cite{witharana2022survey}.

To advance functional verification techniques, it is essential to explore and comprehend the various methodologies employed in this domain. These include static verification methods, often referred to as formal verification \cite{grimm2018survey, grosse2013enhanced, drechsler2010formal, riener2012faust, fey2011assessing}, dynamic approaches, commonly known as simulation-based verification techniques \cite{RostamiIOLTS, loghi2005dynamic, heidari2024enhancing}, and semi-formal strategies, widely recognized as assertion-based verification methods \cite{witharana2022survey, datta2004assertion, li2009study, alatoun2021soc}. Formal verification focuses on ensuring functional correctness through rigorous proofs rather than simulations \cite{drechsler2004advanced}. This category encompasses techniques like model checking, theorem proving, and equivalence checking \cite{drechsler2004advanced}. On the other hand, simulation-based verification evaluates design behavior through simulations, verifying its correctness within the simulated environment \cite{loghi2005dynamic, 1337670, 10.1007/11757283_1}. While each verification approach offers distinct advantages, they also face challenges, such as scalability issues in formal methods and a lack of exhaustiveness in simulation-based techniques \cite{drechsler2004advanced, Zhaolv, articlesimbased}.


While formal and simulation-based verification techniques show significant potential, assertion-based verification has gained traction as a widely adopted method for verifying complex digital systems. ABV leverages the advantages of both formal and simulation-based approaches to ensure design correctness \cite{sohofi2014assertion, lammermann2010towards, ATeam, germiniani2023complete, 10.1007/978-3-642-41010-9_6, 1595644, 7938994}. This technique can be applied at multiple levels of verification abstraction, ranging from high-level assertions embedded in transaction-level testbenches to low-level assertions synthesized directly into hardware implementations \cite{witharana2022survey, sohofi2014assertion, GHAREHBAGHI2007269, Niemann2006AssertionBasedVO, 1656859}. The effectiveness of assertion-based verification primarily depends on the quality of the assertions themselves. These assertions are regarded as valid and accurate rules that describe the behavior and functionality of designs, which the designs must adhere to \cite{ATeam, RezaDFT, Iman_2024}.

Assertion-based verification integrates elements of both formal verification and simulation-based verification, providing a more efficient and effective method for verifying complex digital systems \cite{witharana2022survey}. This approach relies on assertions, which are Boolean expressions that define the expected behavior of a design and must remain true throughout its execution \cite{BZ08}. Within the scope of assertion-based verification, assertions can be defined through two main methods: manual assertion generation and automated assertion mining \cite{heidari2024enhancing, HeidariIman2025MinimizationChapt, 10.1007/978-3-319-46097-0_10}.


Manual definition of assertions demands substantial human expertise and a deep understanding of the design's functionality \cite{BZ08, ARTmine10546742, heidari2024enhancing, Fey2008prop}. This process is often costly, time-intensive, and prone to errors \cite{ARTmine10546742, KGG99, HeidariIman2025chaptsec}. To address these challenges, significant efforts have been made to automate assertion mining \cite{ATeam, hertz2013mining, Vasudevan2010, Malburg2017, 8876892, 5783129, inproceedingsfeyjanmalburg, DBLP:journals/iet-cdt/RoginKFDR09, 10.1145/1403375.1403506, 9599865, harm, exharm, Vasudevan2019goligoldmine, Sheridan2011GoldMineA23}. These automated tools employ various techniques, such as constructing Finite State Machines (FSMs) \cite{ATeam} or leveraging data mining algorithms \cite{iman2023anomalous, ADAssure10546519, 10.1007/978-3-031-70595-3_30}, to generate assertions without human intervention.
Despite their potential, automated assertion miners still face several limitations. One major issue is their inability to produce assertion sets that comprehensively cover all aspects of design behavior, including corner cases. This lack of comprehensive coverage can lead to assertion sets that fail to capture the complete range of design behaviors. Additional shortcomings of current assertion miners include the generation of excessive assertions, redundancy (\textit{i.e.,} assertions that describe the same design behavior), inconsistency (\textit{i.e.,} assertions that conflict with one another), lengthy execution times for assertion mining, and lack of readability of the generated assertions \cite{rezadsd, dominance}. These issues collectively increase the risk of making the verification process more error-prone, time-consuming, and expensive.

The automatic mining of assertions for digital designs has been extensively explored in the literature. The study in \cite{Malburg2017} introduced an assertion miner that uses dynamic dependency graphs to identify signal relationships within a design and generate corresponding assertions. In \cite{8876892}, a syntax-guided enumeration approach was proposed for assertion mining. The technique presented in \cite{ATeam} leverages predefined templates in the form of finite state machines to extract assertions. HARM 
is a hint-based assertion miner that utilizes simulation traces to generate Linear Temporal Logic (LTL) assertions \cite{harm}. The extended version of HARM incorporates a clustering algorithm to refine the mined assertion sets \cite{exharm}.
GoldMine, introduced in \cite{Vasudevan2010}, employs formal verification and static code analysis to derive assertions for Register-Transfer Level (RTL) designs. The method proposed in \cite{9599865} focuses on extracting assertions by converting sequential designs into pseudo-combinational designs. Additionally, ARTmine, a recent advancement in automatic assertion mining, employs specifically designed data mining techniques and algorithms to generate assertion sets \cite{ARTmine10546742}.


Therefore, this manuscript builds upon our previously published survey \cite{11006673}, providing a revised and streamlined presentation focused on conceptual foundations and methodologies in automatic assertion miners. The main contributions are summarized as follows:
\begin{itemize}
    \item We present an analytical study and comparison of the most well-known and recent automatic assertion miners. 

    \item This study summarizes the strengths and limitations of state-of-the-art automatic assertion miners, providing insights for researchers and engineers in the field of verification to better enhance future generations of assertion miners by addressing their shortcomings and to aid in selecting the most appropriate miners for their work.
\end{itemize}
\par The paper is organized as follows: Section~\ref{sec:related works} provides a description of the most advanced automatic assertion miners, 
and Section~\ref{sec:conclusion} concludes the paper.
\section{Automatic Assertion Miners}
\label{sec:related works}
In this section, 
we provide detailed descriptions of the most commonly used and well-known automatic assertion miners by researchers in the field. The goal is to highlight their strengths and shortcomings, offering insights into the limitations of current assertion miners and paving the way for addressing these issues in future generations of automatic assertion miners. The strengths and limitations discussed in this paper are not solely derived from the existing literature on assertion miners, but also from our hands-on experience working directly and indirectly with these tools and their generated assertions over several years of research in this field.

\subsection{GoldMine \cite{Vasudevan2010}}
GoldMine is an automated tool designed to generate assertions for RTL designs. Its methodology involves generating simulation data, mining assertions using a decision tree algorithm, and verifying them formally. It employs static analysis techniques, such as cone-of-influence analysis, to narrow the focus to causal variables, improving the relevance of generated assertions. GoldMine adapts its approach through various configuration options like sampling strategies and reset signal frequency to handle combinational and sequential modules. The tool was evaluated on the Rigel RTL design, demonstrating its capability to capture both simple and complex design behaviors, including temporal relationships across multiple logic levels. In the following, we briefly provide the key strengths and limitations of GoldMine:

\textbf{Key Strengths:}
\begin{itemize}
    \item Automation: GoldMine reduces manual effort by automating the assertion generation process, covering large design spaces efficiently.
    \item Complexity Handling: It captures deep temporal and combinational relationships, providing insights into unintentional but valid design behaviors.
    \item Design Coverage: Its assertions uncover unexplored regions of the design, contributing to improved regression test quality.
\end{itemize}

\textbf{Limitations and Areas for Improvement:}
\begin{itemize}
    \item Dependence on Simulation Data: The quality of assertions heavily relies on diverse and high-quality simulation data, which may not always be available.
    \item Heuristic Reliance: When decision-making in the assertion mining process is inconclusive, heuristics are used, which can lead to trivial or irrelevant assertions.
    \item Handling Temporal Properties: Sequential behaviors require significant methodological fine-tuning, limiting its out-of-the-box applicability for complex temporal relationships.
    \item Over-Constrained Assertions: Some assertions, while correct, may be too trivial or restrictive to add practical value.
    \item Limited Support for Non-Verilog/RTL Designs: GoldMine is limited to working with Verilog designs and does not support SystemVerilog. Additionally, it can only mine assertions from designs in RTL format, meaning it cannot mine assertions for hardware designs implemented in the gate-level format.
\end{itemize}
In conclusion, GoldMine’s ability to automate assertion generation and uncover unexplored design behaviors makes it a valuable tool for early-stage RTL verification. However, future research should address its limitations, particularly in improving the handling of temporal complexity, reducing dependence on simulation data, and refining its decision-making processes, to make it more robust and efficient.

\subsection{DDG \cite{Malburg2017}}
This assertion miner leverages Dynamic Dependency Graphs (DDGs) to describe the relationships between signals in a hardware design and automatically generate properties. The methodology involves first generating DDGs from a set of use cases, followed by extracting initial properties from these graphs. After applying refinement and abstraction steps, general properties are created, which are then verified using a model-checker. This approach aims to automatically identify relevant properties with minimal use case data. In the following, we briefly provide the key strengths and limitations of DDG:

\textbf{Key Strengths:}
\begin{itemize}
    \item Efficient Use Case Requirements: The method is effective even with minimal simulation data. As few as one use case can be sufficient to generate useful properties, which contrasts with other methods that may require extensive simulation runs.
    \item Word-Level Properties: By utilizing DDGs, the approach can generate detailed, word-level properties using a variety of operators. This offers more flexibility and precision in capturing signal relationships.
    \item Generalization of Properties: The refinement and abstraction steps enable the creation of generalized properties that are not tied to specific simulation data, making them broadly applicable across different design scenarios.
    \item Redundancy and Inconsistency: Our limited access to the generated assertions of this miner for some benchmarks suggests that DDG is one of the best at minimizing redundant and inconsistent assertions.
\end{itemize}
\textbf{Limitations and Areas for Improvement:}
\begin{itemize}
    \item Scalability: As the design size increases, the complexity of extracting and processing properties from DDGs may lead to longer processing times, potentially making the method less efficient for large designs.
    \item Internal State Exploration: The method primarily focuses on observable output signals and their relationships, which may limit its ability to generate useful properties for internal signal states. Expanding the scope to include these internal states would improve its applicability for more complex designs. Our limited access to the generated assertions for specific benchmarks showed that the design behavior coverage provided by this miner is considerably lower compared to other miners.
    \item Model-Checker Bottleneck: The reliance on model-checking to verify properties may introduce a bottleneck, especially when dealing with large designs or a high volume of generated properties. 
\end{itemize}
In conclusion, the property mining approach using Dynamic Dependency Graphs offers an efficient and flexible way to generate useful properties with minimal simulation data. While it excels in creating word-level properties and generalizing across use cases, challenges remain in scalability, internal state analysis, and verification speed. Addressing these challenges will enhance the approach's utility for a broader range of hardware verification tasks.

\subsection{A-TEAM \cite{ATeam}}
A-TEAM is an assertion mining tool designed to extract assertions from execution traces of the design under verification across various abstraction levels, including TLM, RTL, and gate-level designs. It is template-driven, allowing users to define custom templates to guide assertion generation, making it more versatile than conventional miners constrained by fixed patterns.
A-TEAM follows a three-step methodology: (1) generate candidate assertions using execution traces and templates, (2) evaluate assertions by injecting testable faults and measuring their fault coverage, and (3) minimize the assertion set using an Integer Linear Programming (ILP)-based approach to retain only assertions that maximize coverage. The result is a concise set of assertions that effectively capture DUV behaviors.
In the following, we briefly provide the key strengths and limitations of A-TEAM:

\textbf{Key Strengths:}
\begin{itemize}
    \item Customizable and Flexible Mining:
The use of user-defined templates enables A-TEAM to explore diverse assertion patterns beyond pre-defined structures, making it adaptable to various abstraction levels.
\item Compact Assertions with High Coverage:
A-TEAM produces fewer but more compact assertions, achieving higher fault coverage compared to decision-tree-based miners.
\item Wide Applicability:
A-TEAM can operate on execution traces from different DUV abstraction levels, increasing its scope of application.
\item Design Behavior Coverage: Our limited access to the generated assertions of this miner for a few benchmarks showed that, although the miner does not offer as high a design behavior coverage as newer generations of miners, it still provides competitive coverage.
\end{itemize}

\textbf{Limitations and Areas for Improvement:}
\begin{itemize}
    \item High Computational Overhead:
The ILP-based minimization and antecedent justification significantly increase execution time, often exceeding that of simpler approaches.
\item Reliance on User-Defined Templates:
The miner's effectiveness depends heavily on the quality of templates, requiring substantial user expertise to achieve optimal results.
\item Limited Temporal Exploration:
While A-TEAM supports multiple temporal templates, its ability to capture complex temporal dependencies could be further enhanced.
\item Scalability Challenges:
The computational demands and heuristic-based decision-making may struggle to scale efficiently for large and complex DUVs.
\end{itemize}
In conclusion, A-TEAM is a versatile and effective assertion mining tool that outperforms traditional approaches in coverage and compactness of mined assertions. However, its reliance on user expertise, high execution time, and limited exploration of complex temporal behaviors present opportunities for future research to enhance its scalability, reduce its computational demands, and increase its automation in assertion generation.

\subsection{HARM \cite{harm}}
HARM (Hint-based Assertion Miner) and its extended version \cite{exharm}, which can be considered an enhanced version of A-TEAM, are robust, template-driven assertion mining tools designed for hardware verification. HARM methodology revolves around leveraging user-defined templates to mine assertions efficiently and flexibly from input traces. By utilizing a three-level parallelization strategy, customizable ranking metrics, and filtering mechanisms, HARM achieves significant computational speed and scalability. Furthermore, the miner allows users to define domain-specific metrics for scoring and ranking assertions, tailoring its outputs to context-specific needs.
In the following, we briefly provide the key strengths and limitations of HARM:

\textbf{Key Strengths:}
\begin{itemize}
    \item Customizable and Flexible Methodology: HARM’s template-based approach allows users to tailor the mining process to specific domains, ensuring the relevance of the generated assertions to the verification context.
    \item Highly Parallelized Algorithms: The three-level parallelization strategy (across evaluation units, template permutations, and templates) fully exploits CPU cores, achieving substantial speed-ups, particularly for large datasets.
    \item User-Defined Context-Aware Metrics: HARM empowers users to define metrics for filtering and ranking assertions, providing fine-grained control over the results. This makes it adaptable to various verification needs and scenarios.
    \item Scalability and Efficiency: Experimental results demonstrate that HARM scales effectively with input size and outperforms state-of-the-art miners in terms of both speed and assertion quality.
    \item Security Context: Our experiments utilizing HARM's assertions in the context of security and Hardware Trojan detection demonstrate its significant applicability in detecting Hardware Trojans.
\end{itemize}

\textbf{Limitations and Areas for Improvement:}
\begin{itemize}
    \item Template Dependency: While the template-based approach provides flexibility, it also introduces a dependency on the user’s ability to define meaningful templates. Ineffective or incomplete templates may result in suboptimal or irrelevant assertions.
    \item Metric Calibration Complexity: The ranking system, while flexible, requires users to select from numerous calibrate function configurations. This complexity may pose challenges for less experienced users and could benefit from automated or guided metric tuning.
    \item Limited Exploration of Unknown Assertions: HARM’s reliance on templates inherently limits its ability to discover assertions outside the predefined template scope, potentially missing novel or unexpected behaviors.
    \item Ranking Bias: The ranking formula gives higher weight to assertions that score well across all metrics but penalizes assertions with low scores in any single metric. While effective for filtering poor-quality assertions, this may undervalue assertions with unique but isolated strengths.
    \item Evaluation Limited to Traces: HARM’s methodology evaluates assertions solely on the given input trace(s). This limits its applicability in scenarios where traces are incomplete or do not fully represent the system’s behavior.
    \item High Number of Generated Assertions: Our experimental analysis reveals that HARM generates a significantly higher number of assertions than all other miners, without necessarily offering better design behavior coverage. This can potentially lead to a lengthier verification process.
\end{itemize}
In conclusion, HARM is a powerful and efficient tool for hint-based assertion mining in hardware verification. Its template-driven methodology, three-level parallelization, and context-aware ranking mechanisms position it as a competitive miner in the field. However, its reliance on user-defined templates, the complexity of metric calibration, and its computational demands highlight areas where further research could improve its usability and performance. Addressing these limitations—such as reducing template dependency, automating metric tuning, and broadening the exploration of non-template assertions—would make HARM more accessible and versatile, paving the way for advancements in assertion mining methodologies.
\subsection{ARTmine \cite{ARTmine10546742}}
ARTmine is an automatic assertion miner designed to generate temporal assertion sets for hardware verification. The core methodology of ARTmine revolves around data mining algorithms designed for mining association rules from simulation traces of hardware designs, with the goal of identifying critical design behaviors. These mined association rules are then transformed into temporal assertions for verification. 
The mining process is designed to be both efficient and effective, with a specific focus on corner cases by adjusting the parameters of the miner. This ensures the tool generates a minimal but highly relevant and accurate set of assertions. ARTmine has demonstrated strong capabilities in mutant detection and assertion generation, particularly when compared to other assertion mining tools like HARM and GoldMine. In the following, we briefly provide the key strengths and limitations of ARTmine:

\textbf{Key Strengths:}
\begin{itemize}
    \item Efficient Assertion Generation: ARTmine is designed to generate a minimal number of assertions that are highly relevant and effective for verification. By focusing on corner cases and using a controlled approach with configurable parameters in the tool, ARTmine avoids the overgeneration of assertions. This is crucial in preventing the verification process from becoming overwhelmed by unnecessary assertions, which is often a limitation in other assertion miners.
    \item Effective Mutant Detection: ARTmine has shown superior performance in detecting injected mutants. This highlights its ability to identify significant design behaviors that may not be captured by traditional assertion mining techniques. The tool’s focus on corner cases further enhances its detection capabilities, making it particularly valuable in the verification of complex hardware systems.
    \item Scalability: ARTmine is capable of effectively handling large-scale designs. The tool can efficiently process simulation traces and generate assertions, even for designs with a large number of components. This scalability is a significant advantage for applying ARTmine to real-world hardware verification tasks.
    \item Support for Temporal Patterns: ARTmine’s support for a variety of temporal patterns, allows it to handle a wide range of verification requirements. This flexibility makes ARTmine applicable to a diverse set of design scenarios, from simple to more complex verification tasks.
    \item Security Context: Our experimental analysis of ARTmine's assertions in the realm of security and Hardware Trojan detection highlights its strong applicability in identifying Hardware Trojans.
\end{itemize}

\textbf{Limitations and Areas for Improvement:}
\begin{itemize}
    \item Limited Temporal Pattern Support: While ARTmine supports several key temporal patterns, its scope is limited to specific predefined patterns. Other temporal patterns commonly found in hardware verification, are not directly supported in the current implementation. Extending ARTmine to handle a wider range of temporal patterns could further enhance its applicability and make it more versatile for various hardware verification scenarios.
    \item Data Mining Thresholds: ARTmine is designed to handle corner cases effectively, but the tool’s reliance on predefined data mining thresholds could lead to the exclusion of some important but rare design behaviors. 

    \item Scalability Challenges for Extremely Large Designs: Although ARTmine demonstrates good scalability with medium-to-large designs, it may face challenges when applied to extremely large designs with significantly more I/O or state variables for one of its predefined temporal templates. As the number of variables increases, the number of mined assertions and their associated computational overhead may grow exponentially. Optimizing the underlying algorithms to better scale with extremely large designs would make ARTmine more applicable to cutting-edge hardware systems, especially in the context of modern System-on-Chips (SoCs) and extremely large designs.
\end{itemize}

In conclusion, ARTmine is a strong and efficient automatic assertion miner that excels in generating temporal assertions for hardware verification. Its ability to handle a variety of temporal patterns, detect mutants with fewer assertions compared to other miners, and scale to large designs makes it a valuable tool for hardware verification tasks. However, it does have certain limitations, such as the restricted set of supported temporal patterns and reliance on predefined thresholds. These limitations present opportunities for future research to extend ARTmine’s capabilities, improve its handling of complex behaviors, and enhance its scalability for next-generation hardware systems. With further improvements in these areas, ARTmine could become a powerful tool for comprehensive and efficient hardware verification across a wide range of design environments.

\section{Conclusion}
\label{sec:conclusion}
In this paper, we surveyed and analyzed the most well-known and widely used automatic assertion miners to better explore their strengths and weaknesses. Our analysis highlighted their pros and cons, emphasizing the current limitations of state-of-the-art automatic assertion miners. By focusing on these limitations, we aimed to guide future improvements in automatic assertion miners, ultimately enhancing the functional verification of hardware designs. \section*{Acknowledgement} This work is partially funded by the “Resilient Trust” project of the EU’s Horizon Europe research and innovation programme under grant agreement No. 101112282.

\bibliographystyle{IEEEtran}
\begingroup
\raggedleft
\bibliography{references/references}
\endgroup

\vspace{12pt}


\end{document}